%% file: bmumuhdraft2_2.tex
\RequirePackage{lineno}
\documentclass[aps,prd,twocolumn,superscriptaddress,showpacs]{revtex4}

\usepackage{graphicx}
\usepackage{dcolumn}
\usepackage{bm}

\begin{document}

\title{Search for the Rare Decays $B^+ \to \mu^+\mu^- K^+$, $B^0\to\mu^+\mu^- K^*(892)^0$, and $B_s^0\to\mu^+\mu^- \phi$ at CDF}

\input{Jan_2008_Authors_Visitors1_oldeman_tesarek}


\begin{linenomath}
\begin{abstract}
We search for $b\to s\mu^+\mu^-$ transitions in
$B$ meson ($B^+$, $B^0$, or $B^0_s$) decays with 924\,pb$^{-1}$ of $p\bar{p}$ collisions at $\sqrt{s}=1.96$\,TeV collected with the CDF II 
detector at the Fermilab Tevatron. We find excesses
with significances of 4.5, 2.9, and 2.4 standard deviations in
the $B^+\to\mu^+\mu^-K^+$,
$B^0\to\mu^+\mu^-K^*(892)^0$, and $B_s^0\to\mu^+\mu^-\phi$ decay modes, respectively.
Using $B\to J/\psi h$ ($h$ = $K^+$, $K^*(892)^0$, $\phi$) decays as normalization channels, we
report branching fractions for the previously observed $B^+$ and
$B^0$ decays, ${\cal B}(B^+\to\mu^+\mu^-K^+)=(0.59\pm0.15\pm0.04)\times10^{-6}$, and ${\cal B}(B^0\to\mu^+\mu^-K^*(892)^0)=(0.81\pm0.30\pm0.10)\times10^{-6}$,
where the first uncertainty is statistical, and the second is systematic.
These measurements are consistent with the
world average results, and are competitive with the best available
measurements.
We set an upper limit on the relative branching fraction 
$\frac
{{\cal B}(B_s^0\to\mu^+\mu^-\phi)}
{{\cal B}(B_s^0\to J/\psi\phi)}
<2.6(2.3)\times10^{-3}$ 
at the 95(90)\% confidence level, 
which is the most stringent to date.

\end{abstract}
\end{linenomath}

\pacs{Valid PACS appear here}

\maketitle

\newpage

The decay of a $b$ quark into an $s$ quark and two muons ($b\to s
\mu^+ \mu^-$) is a flavor-changing neutral current process, forbidden
at tree level in the standard model (SM) but allowed 
through highly suppressed internal loops. New
physics could manifest itself in a larger branching fraction, a modified
 dimuon mass distribution, or 
angular distributions of the decay products
different from that predicted by the SM
~\cite{Colangelo:2006vm,Kruger:2005ep,Aliev:2003gp,Demir:2002cj}.
In this paper, we report branching ratio measurements of exclusive decays,
where the $s$ quark hadronizes into a single meson.
Reconstructing a final state 
with only three or four charged final state particles results 
in smaller backgrounds than those expected in a search for inclusive $B\to X_s\mu^+\mu^-$ decays.
The rare decays, $B^+\to \mu^+\mu^-K^+$ and $B^0\to
\mu^+\mu^-K^*(892)^0$ have been observed at the $B$
factories~\cite{Aubert:2006vb,Ishikawa:2003cp}, with branching
fractions of $O(10^{-6})$, consistent with SM predictions~\cite{Burdman:1995ks,Colangelo:1995jv,Aliev:1996hb,Melikhov:1997wp,Ali:1999mm,Choi:2001hg,Ali:2002jg,Chen:2002zk}.
The analogous decay in the $B_s^0$ system, $B_s^0\to \mu^+\mu^-\phi$, has a
predicted branching ratio of $1.6\times10^{-6}$~\cite{Geng:2003su},
but has not yet been observed~\cite{Abazov:2006qm,CDFRUN1_bsmumuphi}.
We report branching ratio measurements from 924\,pb$^{-1}$ of 
Collider Detector at Fermilab (CDF) Run II data
that pave the way for future studies of larger datasets from which
kinematic distributions may be measured.  We search for $B\to\mu^+\mu^-h$ decays, where
$B$ stands for $B^+$, $B^0$, or $B_s^0$, and $h$ stands for $K^+$,
$K^*(892)^0$, or $\phi$, respectively.  The $K^*(892)^0$, referred to
as $K^{*0}$ throughout this paper, is reconstructed in the 
$K^{*0}\to K^+\pi^-$, decay mode,
and the $\phi$ meson is reconstructed as $\phi\to
K^+K^-$.  We measure branching ratios relative to $B\to
J/\psi h$ decays, followed by $J/\psi\to\mu^+\mu^-$ decays, 
resulting in the same final state particles as the rare decay modes.
Many systematic uncertainties cancel in the relative
branching ratios:
\begin{linenomath}
\begin{equation}
\frac{{\cal B}(B\to\mu^+\mu^- h)}{{\cal B}(B\to J/\psi h)}=\frac{N_{\mu^+\mu^- h}}{N_{J/\psi h}} \frac{\epsilon_{J/\psi h}}{\epsilon_{\mu^+\mu^- h}}\times {\cal B}(J/\psi\to\mu^+\mu^-),
\label{eqn:br}
\end{equation}
\end{linenomath}
where $N_{\mu^+\mu^- h}$ is the observed number of $B\to\mu^+\mu^- h$
decays, $N_{J/\psi h}$ is the observed number of $B\to J/\psi h$
decays, while $\epsilon_{J/\psi h}$ and $\epsilon_{\mu^+\mu^- h}$ are
the combined selection efficiency and acceptance of the experiment for $B\to J/\psi h$
and $B\to\mu^+\mu^- h$ respectively.  Throughout this report, charge
conjugate modes are implicitly included. 

CDF II is a general purpose detector,
 located at the Tevatron
$p\bar{p}$ collider~\cite{Acosta:2003qr}.  Charged particle trajectories (tracks) are
detected by the tracking system comprised of a seven-layer double-sided silicon
microstrip detector and a drift chamber, both in a 1.4\,T axial magnetic field.
The silicon detector~\cite{Sill:2000zz}
 ranges in radius from 1.3 to 28\,cm,
and has a single-hit resolution of
approximately 15\,$\mu$m. The  drift chamber~\cite{Affolder:2003ep}
 provides up to 96
measurements from radii of 40 to 137\,cm with a single-hit resolution of 
approximately 180\,$\mu$m.
  A muon chamber identification
system of plastic scintillators and drift chambers~\cite{Ascoli:1987av} is located on the exterior of the detector with a central part
covering $|\eta|<0.6$, and an extended region covering
$0.6<|\eta|<1.0$, where $\eta$ is the pseudorapidity~\cite{eta}.
In the central(extended) region, muons
are detected if their transverse momentum component, $p_T$,
is greater than 1.5(2.0)\,GeV/$c$.
Events are selected with a three-level trigger system.  The first
trigger level requires the presence of two charged particles
with $p_T\geq 1.5$\,GeV/$c$ ($|\eta|\leq 0.6$)
or  $p_T\geq 2.0$\,GeV/$c$ ($0.6\leq|\eta|\leq1.0$), matched to
track segments in the muon chambers to form muon candidates.  At the second
level, a more restrictive selection is made by requiring that the muon
candidates have opposite charge, and that their opening angle in the
plane transverse to the beamline is less than 120$^\circ$.
At the third trigger level, the event is fully reconstructed.
The trajectories of the muon candidates in the silicon detector
are required to intersect at a 
point which is displaced transversely from the beamline by at
least 100\,$\mu$m.

The offline selection begins with a suppression of
random combinations of tracks satisfying the selection requirements (combinatoric background),
by requiring that all tracks have
$p_T>0.4$\,GeV/$c$ and match hits from at least three layers of the
silicon detector.  
The trajectories of a pair of muon candidates that satisfy the trigger requirements,
and the tracks that form the hadron candidate, 
are fitted with the constraint that they originate from a single vertex in 3-dimensional space to 
form a $B$ candidate.
The $\chi^2$ probability of the fit is required to be greater than $10^{-3}$.
  A $K^+$ candidate is a track assigned the charged kaon mass, and
the $K^{*0}$ ($\phi$) candidates are formed from oppositely charged
pairs of tracks whose invariant mass must lie within 50
(10)\,MeV/$c^2$ of the $K^{*0}$ ($\phi$) mass. 
For all particles, we use the world average values tabulated in Ref.~\cite{Yao:2006px}.
The ambiguity of the mass assignment in
$B^0\to\mu^+\mu^-K^{*0}$ decays is handled by choosing the combination
whose $K^+\pi^-$ mass is closer to the $K^{*0}$ mass.  
  In reconstructing the $B$ candidates, the 
meson containing a strange quark, $h$, is required to have $p_T(h)\geq1.0$\,GeV/$c$, and the $B$
candidate is required to have $p_T(B)\geq4.0$\,GeV/$c$.
We require that the distance of closest approach between the 
flight path of the $B$ candidate and the beamline,
$|d_0(B)|$, is less than $120\,\mu$m, to reduce the combinatoric background with
no loss of signal.  

Normalization mode candidates are identified by 
having a dimuon invariant mass within 50\,MeV/$c^2$ of the $J/\psi$
mass, yielding  approximately 12 000 $B^+\to J/\psi K^+$, 4400 $B^0\to J/\psi K^{*0}$,
and 800 $B_s^0\to J/\psi\phi$ decays.
The $B$ candidate mass distributions of all modes are compatible with a Gaussian of width $\sigma=20$\,MeV/$c^2$.
To reduce backgrounds from $B$ decays to mesons containing $c$ quarks,
several vetoes are applied to $B\to\mu^+\mu^-h$ candidates, listed below.
We eliminate candidates with a dimuon
mass near the $J/\psi^($$'$$^)$ : $2.9\leq
m_{\mu\mu}\leq3.2$\,GeV/$c^2$ or $3.6\leq
m_{\mu\mu}\leq3.75$\,GeV/$c^2$ respectively.
$B\to J/\psi^($$'$$^) h$ decays followed
by the radiative decay of the $J/\psi^($$'$$^)$ into two muons and a photon
that is not reconstructed,
may have a dimuon mass that passes the $J/\psi^($$'$$^)$ veto.
We reject these events making use  of the correlation between the candidate's invariant mass
and the dimuon invariant mass, and require $|(m_{\mu\mu h}-m_B)-(m_{\mu\mu}-m_{J/\psi(')})|>100$\,MeV/$c^2$.
$B\to J/\psi^($$'$$^) h$ decays with one hadron misidentified as a muon
form a potential background to the rare decay search. 
We reject this class of background by requiring that all combinations
of tracks of the candidate have an invariant mass that differs
by at least 40\,MeV/$c^2$ from the $J/\psi^($$'$$^)$ mass.
We also reject candidates with track pairs with a mass within $\pm25$\,MeV/$c^2$ of the $D^0\to K^-\pi^+$ decay,
or track triplets compatible with a mass within $\pm25$\,MeV/$c^2$ of the
$D^+\to K^-\pi^+\pi^+$ or $D_s^+\to K^+K^-\pi^+$ decays.

 We further improve the signal to background ratio
by an optimization of the selection based on discriminating variables.  For this purpose, 
Monte Carlo (MC) simulations of the signal processes are used.
We generate single $b$ hadrons using the transverse momentum spectrum from $B\to J/\psi X$, measured by CDF~\cite{Acosta:2003qr}.
Decays of all $b$ hadrons are simulated using {\sc evtgen}~\cite{Lange:2001uf}, with lifetimes
 from Ref.~\cite{Yao:2006px}, allowing for a decay width
difference between the $B_s^0$ mass eigenstates,
$\Delta\Gamma/\Gamma=0.12\pm0.06$~\cite{deltagamma}.
The decays
$B^0\to J/\psi K^{*0}$ and $B_s^0\to J/\psi\phi$ are simulated according to
the polarization amplitudes measured by CDF~\cite{cdfdeltagamma}.  The
dynamics of the rare decay processes are simulated according to the calculations of
Ali {\it et al.}~\cite{Ali:1999mm}.  For $B_s^0\to\mu^+\mu^-\phi$, we assume a
mixture of 50\% CP-even and 50\% CP-odd states.  The
interactions of final state particles are simulated 
using a {\sc geant}~\cite{GEANT} model of the CDF II detector, digitized into the
CDF event format, and reconstructed using the same software
as in the processing of collision data.
The detector simulation includes a full emulation of the CDF trigger system.
We find that the transverse momentum spectrum of the $B$ mesons
of this measurement are on average higher than in the simulation.
This can be explained by the presence of a $B$ baryon component with
low transverse momentum in the measurement
used as input spectrum of the simulation~\cite{Acosta:2003qr}.
To correct for this, the simulated events are weighted by a
polynomial function, which is obtained from the ratio of data to
MC transverse momentum distribution of the $B$ meson.  

\begin{figure*}
\includegraphics[width=5.5cm]{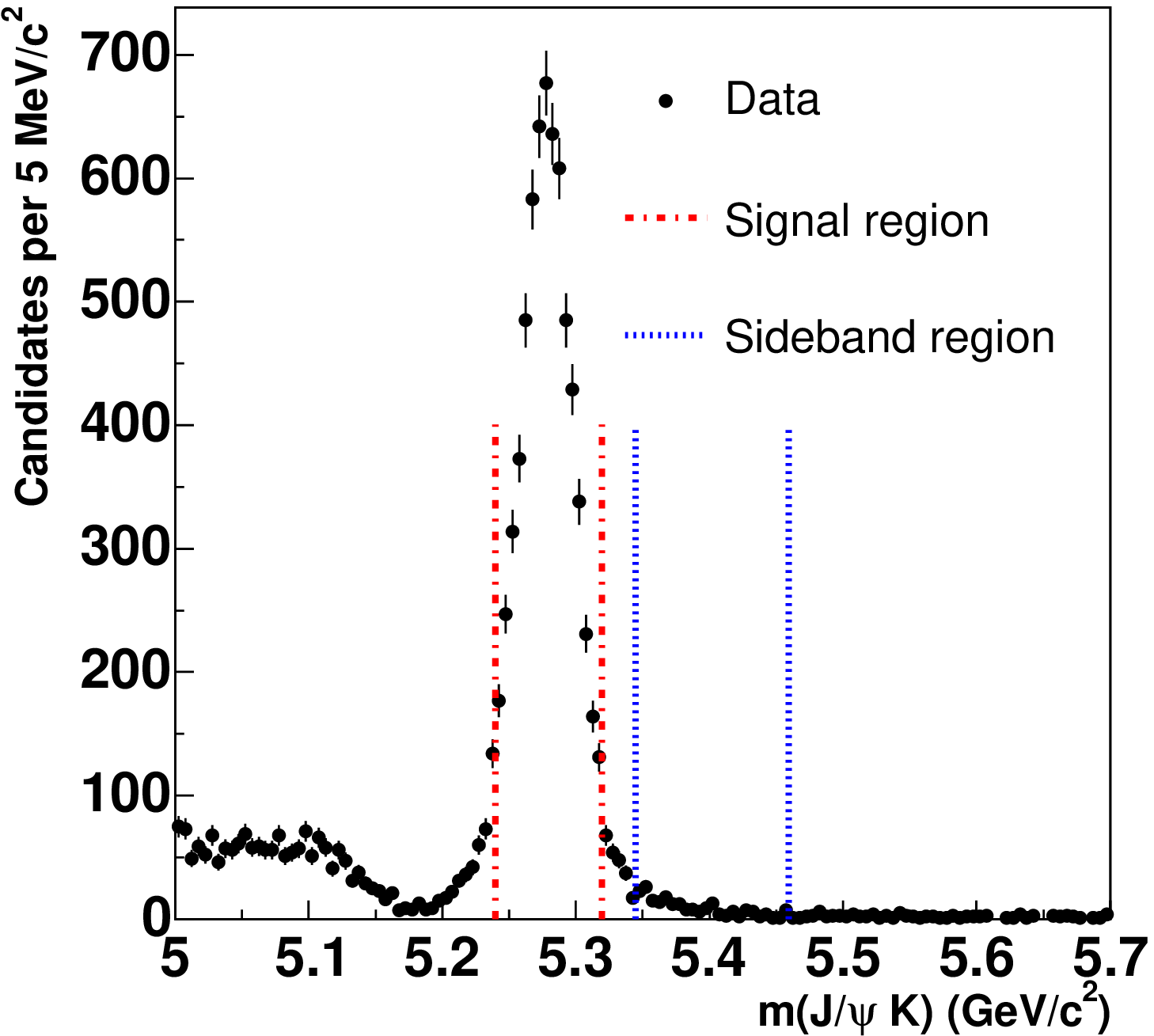}\put(-18,126){a}
\includegraphics[width=5.5cm]{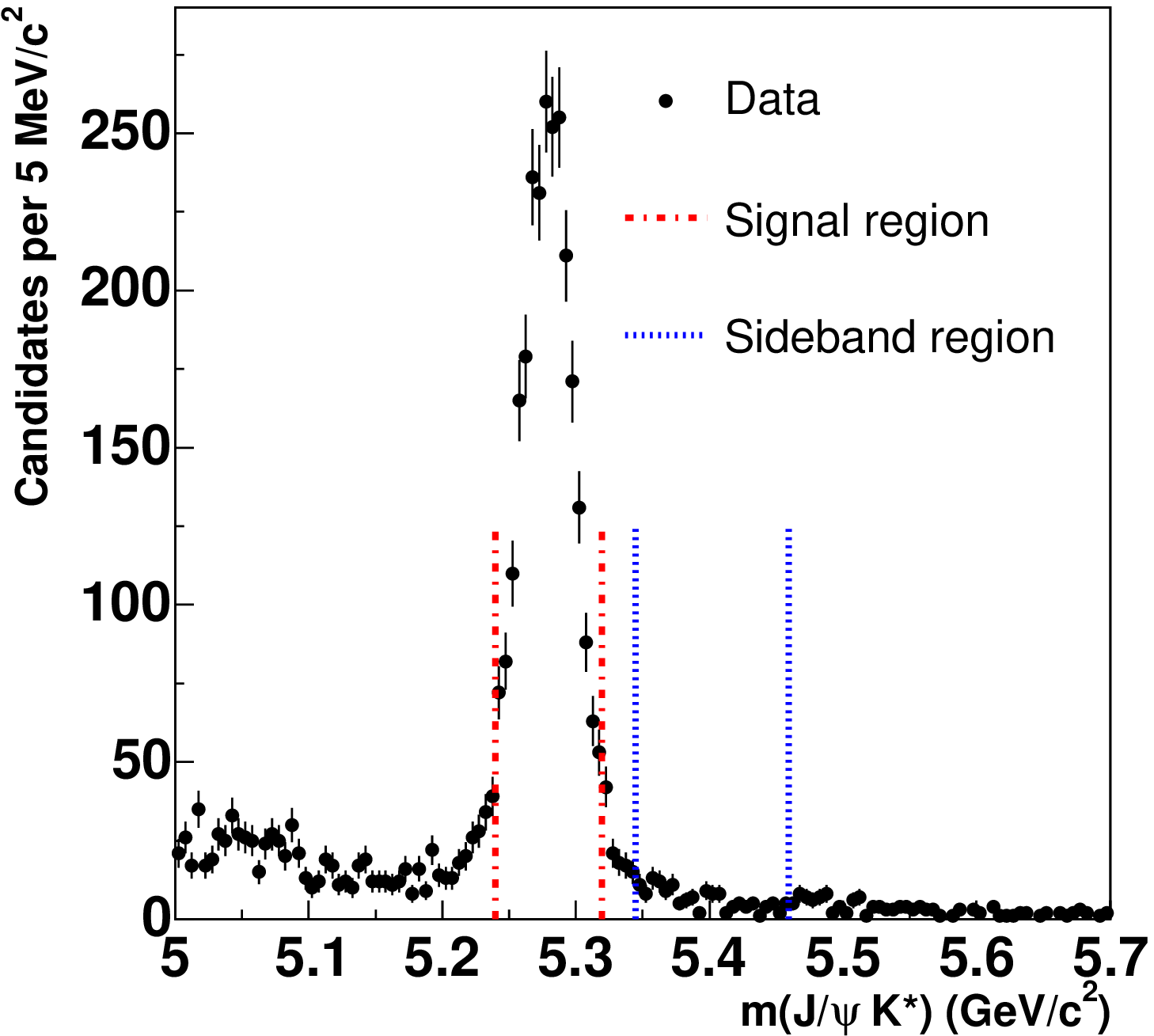}\put(-18,126){b}
\includegraphics[width=5.5cm]{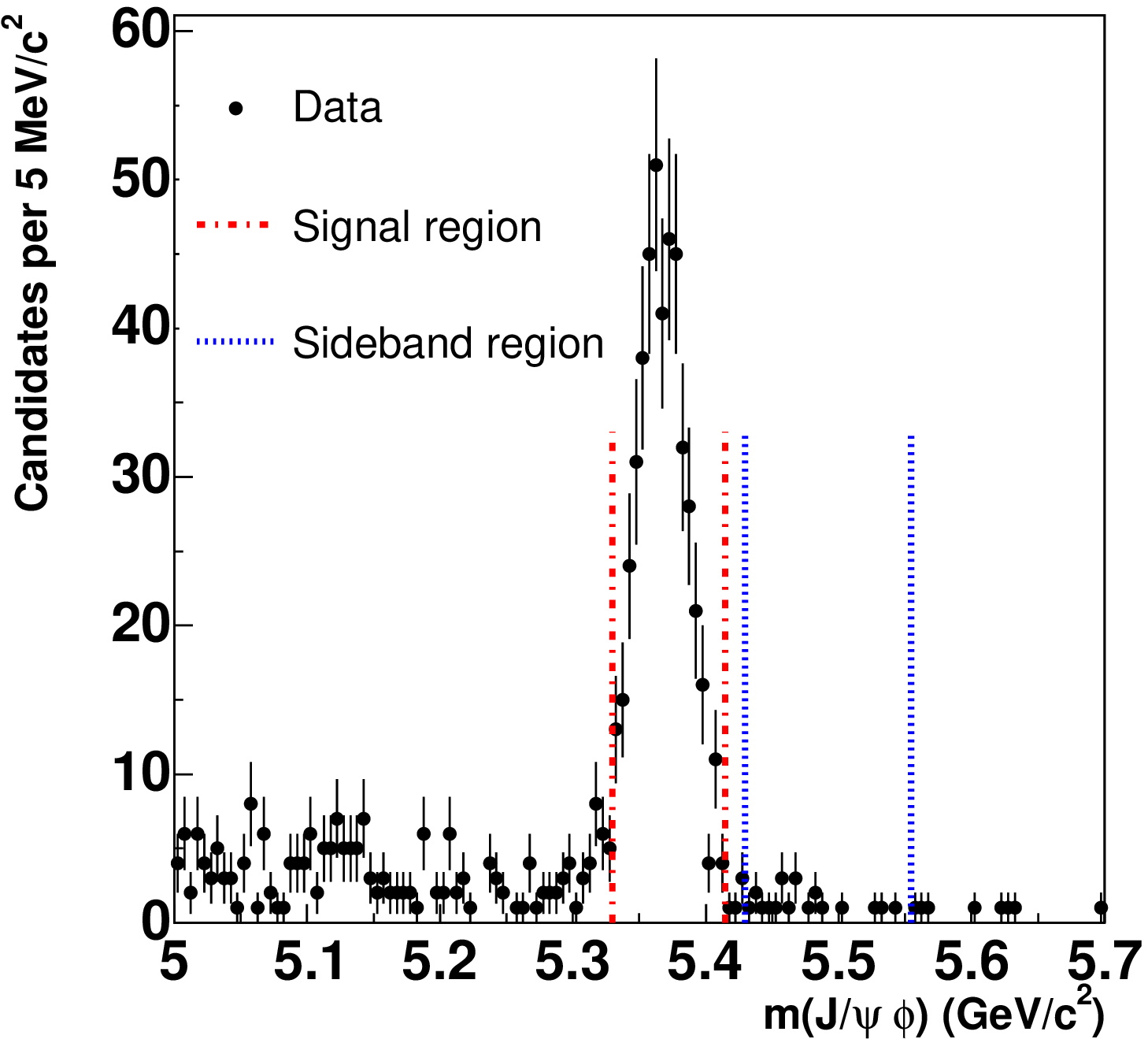}\put(-18,126){c}
\caption{Invariant mass spectra of (a) $J/\psi K^+$, (b) $J/\psi K^{*0}$ and 
(c) $J/\psi \phi$ candidates after applying optimized selection requirements.}  
\label{fig:jpsitight}
\end{figure*}

\begin{figure*}
\includegraphics[width=5.5cm]{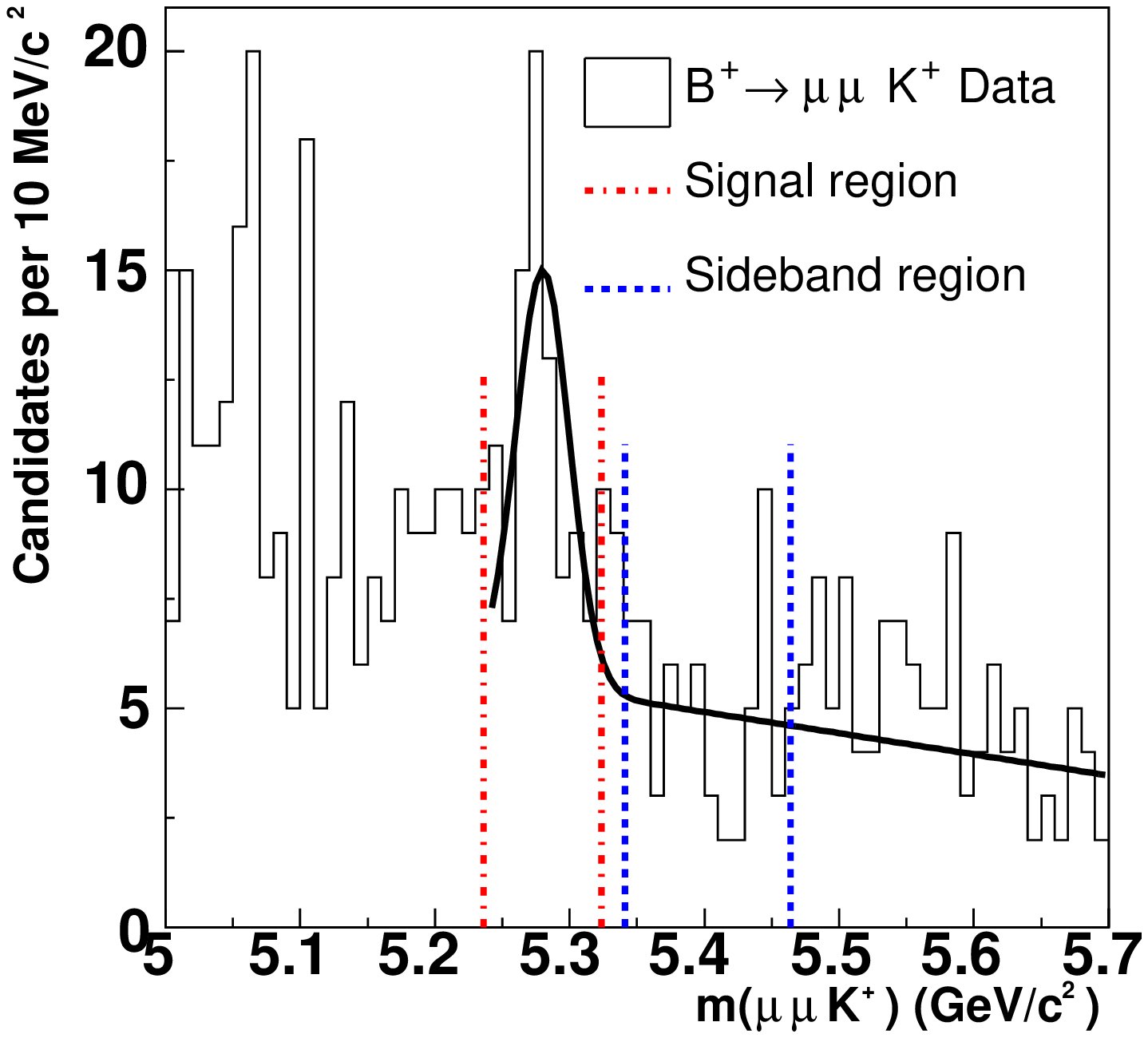}\put(-110,126){a}
\includegraphics[width=5.5cm]{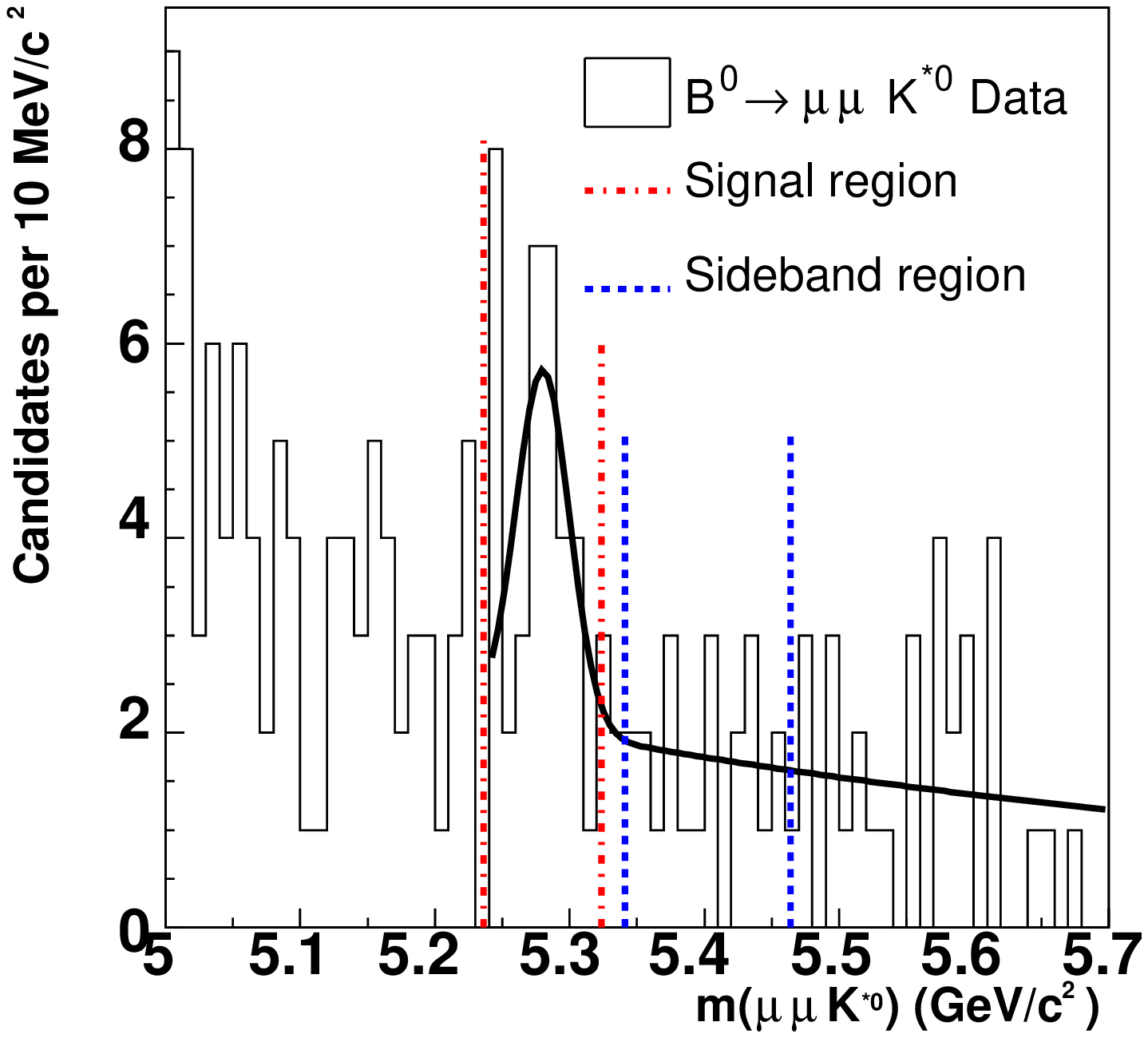}\put(-110,126){b}
\includegraphics[width=5.5cm]{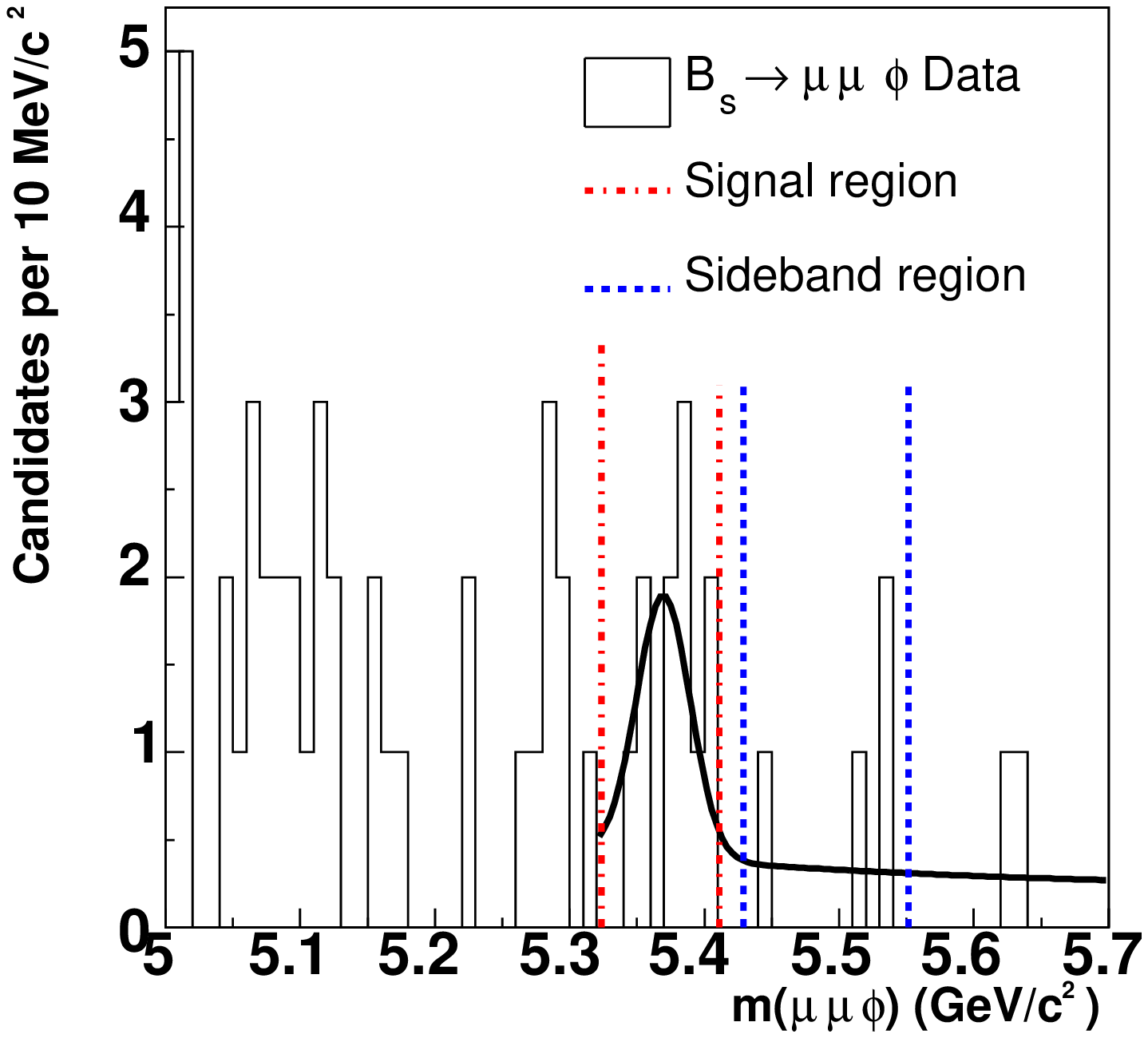}\put(-110,126){c}
\caption{Invariant mass spectra of (a) $\mu^+\mu^- K^+$, (b) $\mu^+\mu^- K^{*0}$ and (c) $\mu^+\mu^- \phi$ candidates.
The superimposed curves are a sum of a single Gaussian with a width of 20\,MeV/$c^2$ representing the signal,
and a curve representing the background as determined by the procedure described in the text.
The curves are not drawn for masses below the signal window, 
since they are not expected to predict the background level where
partially reconstructed $B$ decays contribute.}
\label{fig:unblind}
\end{figure*}

We tighten the candidate selection to obtain the smallest
expected statistical uncertainty on the measurement of the $B\to\mu^+\mu^-h$ branching
ratios by finding the selection with the largest value of
$N_{sig}/\sqrt{N_{sig}+N_{bkg}}$.  The expected number of signal events, $N_{sig}$, is determined by scaling the MC
yields of the rare decay modes to the yields of the normalization channels in the
data, and correcting by the relative branching ratios and selection
efficiencies. 
This procedure requires estimates of the branching ratios of the rare decay modes. 
For the $B^+ \to \mu^+\mu^- K^+$ and $B^0\to\mu^+\mu^- K^{*0}$ decays we use
the world average measured branching ratios~\cite{Yao:2006px},
while for $B_s^0\to\mu^+\mu^- \phi$ we use the theoretical estimate~\cite{Geng:2003su}.
The expected number of background events, $N_{bkg}$, is estimated by extrapolating the
candidate yield with an invariant mass $(180-300)$\,MeV/$c^2$ higher than the $B$ mass,
to the signal region.
Candidates with an invariant mass lower than the $B$ mass are 
not suitable for background estimates, since they contain partially reconstructed $B$ decays,
not expected in the signal region.
We find good discriminating power between signal and
background from the following three quantities:  $t/\sigma_{t}$, $\alpha$, and $I$.
The significance of the proper decay time, $t/\sigma_{t}$,
is defined as the proper decay time of the $B$ candidate, divided by its uncertainty.
The angle $\alpha$ is defined as the difference in angle between the $B$
candidate's momentum vector and the vector from the primary vertex to the
$\mu\mu h$ vertex. The isolation, $I$, is defined as
the transverse momentum carried by the $B$ meson candidate divided by
the transverse momentum of all tracks in a cone of $\Delta R=\sqrt{\Delta \eta ^2 + \Delta \phi^2}=1.0$ around the
direction of the $B$ meson candidate, including those of the $B$
candidate itself.  Here $\Delta\eta$ is the
difference in pseudorapidity of the $B$ candidate and each track, and
$\Delta\phi$ is the difference in their azimuthal angles.  Scanning
 different combinations of selection thresholds, we find that
the optimal values are very similar for the three rare decay modes:
$t/\sigma_{t}\geq14$, $\alpha\leq 60$\,mrad,
and $I\geq0.6$.  Applying the optimized selection requirements
to the normalization channels yields the distributions shown in
Fig.~\ref{fig:jpsitight}. The shape of the combinatoric
background is estimated from a sample with poor vertex quality
($\chi^2$ probability $<10^{-3}$).
The invariant mass distribution of this poor vertex quality sample is fitted 
with a Gaussian distribution for the remaining signal contribution and an 
exponential plus a constant to model the background.
We estimate the background in the signal region of the tight selection
using this functional form, normalized to the number of candidates 
with invariant mass $(60-180)$\,MeV/$c^2$ higher than the $B$ mass.
  Using this method, we find $6361\pm82$ $B^+ \to J/\psi K^+$, 
$2423\pm52$
$B^0\to J/\psi K^{*0}$, and $431\pm22$ $B_s^0\to J/\psi \phi$
decays.
  The invariant mass distributions for the rare decay modes are
shown in Fig.~\ref{fig:unblind}.  
In the $\pm40$\,MeV/$c^2$ window around the $B$ mass we find 
90 $B^+ \to \mu^+\mu^-K^+$,
35 $B^0\to\mu^+\mu^- K^{*0}$,
and 11 $B_s^0\to\mu^+\mu^- \phi$ candidates.
From simulation studies we verified that the present selection accepts
decays for every kinematically possible dimuon invariant mass, 
with an efficiency difference not exceeding a factor two,
except for the windows around the $J/\psi^($$'$$^)$ that have been explicitly 
excluded.

For the signal modes, we evaluate three sources of background: the combinatoric background,
hadrons misidentified as muons, and hadrons with misassigned mass.
As in the normalization mode, the shape of the combinatoric
background is estimated from a sample with poor vertex quality
($\chi^2$ probability $<10^{-3}$).
The invariant mass distribution of these candidates is fitted to an 
exponential plus a constant.
This functional form is used to extrapolate the number of candidates 
with invariant mass $(60-180)$\,MeV/$c^2$ higher than the $B$ mass to the signal region.
The resulting backgrounds, and their statistical uncertainties, consist of
$44.3\pm5.8$, $16.3\pm3.6$, and $3.1\pm1.5$ candidates for the $B^+ \to \mu^+\mu^-
K^+$, $B^0\to\mu^+\mu^- K^{*0}$, and $B_s^0\to\mu^+\mu^- \phi$ decay modes,
respectively.  Backgrounds from hadrons misidentified as muons are
estimated from simulation.  Final state particles from simulated charmless $B$
decays are weighted with the muon misidentification probabilities.  The misidentification
probabilities for charged kaons and pions are measured from a sample of
$D^{*+}\to D^0 \pi^+$ followed by $D^0\to K^-\pi^+$ decays.  We estimate a net background of
1.0 events for the $B^+\to\mu^+\mu^-K^+$ channel, primarily coming from misidentified $B^+\to
K^+\pi^-\pi^+$ decays.  For the $B^0$ and $B_s^0$ modes, this background
is negligible.  
From simulation we determine the background contribution that arises from assigning the
wrong masses to the hadrons.  We estimate 0.4 $B^0\to\mu^+\mu^-K^{*0}$
decays which are reconstructed as $B_s^0\to\mu^+\mu^-\phi$ and 0.2
$B_s^0\to\mu^+\mu^-\phi$ decays reconstructed as
$B^0\to\mu^+\mu^-K^{*0}$.

We calculate the signal yield by subtracting the predicted background from the number of candidates in the signal window.
We find an excess in the $B$ signal region in all three channels, and determine the significance by
calculating the Poisson probability for the background to
fluctuate to the number of observed events or higher, taking into
account the uncertainty on the background.
We find an equivalent Gaussian significance of
4.5, 2.9, and 2.4 standard deviations, respectively for the $B^+$,
$B^0$, and $B_s^0$ modes.

Table~\ref{tab:sys} lists the systematic uncertainties
associated with the relative selection efficiency,
discussed in further detail below.
To take into account the uncertainty on the relative efficiency due to uncertainties in
the dynamics of the rare decays, we apply the weak form factors 
from Ref.~\cite{Melikhov:2000yu,Colangelo:1994qr}
and evaluate the largest difference between the reconstruction efficiency
and the central value.
We find an uncertainty of 3.1\% or less.
The uncertainty related to the
$p_T(B)$ spectrum is evaluated from the change in relative efficiency
when the three parameters in the $p_T(B)$ weighting function are
varied by one standard deviation of their values determined from fits to data,
taking into account correlations.
The difference reaches 1.4\% in the $B_s^0$ mode.  The muon trigger
efficiency close to the 1.5\,GeV/$c$ $p_T$ threshold is poorly known.
We find a change in the relative efficiency reaching 1.3\% if the
minimum $p_T$ is varied by $\pm100$\,MeV/$c$, covering the range over
which the trigger efficiency rises from zero to one.
The final state particles of the rare decay modes have approximately 10\% more low momentum
tracks than those of the normalization channels. The simulation models the
track reconstruction efficiency  to an accuracy of
2\% in the $p_T$ range of $0.4-1.5$\,GeV/$c$.
We therefore assign a systematic uncertainty on the
relative efficiency of 2\%$\times$10\%=0.2\%.  We find an uncertainty
of $8.3\%$ on the relative efficiency of $B_s^0$ decays due to the
unknown fraction of the short-lived CP-even state in the rare decay mode.
We assume that the CP-even fraction is 0.5, and assess systematic uncertainties for
the extremes of the fraction valued at 1.0 and 0.0.  In addition, the
uncertainty on $\Delta\Gamma/\Gamma$ contributes another
$2.6\%$, resulting in a total uncertainty of $8.7\%$ associated with
the $B_s^0$ decay width difference.  This is the largest systematic uncertainty
on the relative efficiency of the $B_s^0$ mode.
We evaluate the effect of the uncertainty on the fraction of $J/\psi$
mesons produced with a longitudinal polarization by varying
the fraction measured at CDF~\cite{cdfdeltagamma} by $\pm1\sigma$.  The effect is 0.6\% for
$B^0\to J/\psi K^{*0}$ and 0.1\% for $B_s^0\to J/\psi\phi$.

The statistical uncertainties of yields  
in the normalization channels are
included as systematic uncertainties, which range from 1.3\% for the $B^+$
channel to 5.1\% for the $B_s^0$ channel.  Cabibbo-suppressed $B^+\to
J/\psi \pi^+$ decays contribute 0.1\% to the yield of the
normalization channel for the $B^+$ channel. 
We introduce the full effect into our estimate of the systematic uncertainty 
without correcting the result.
The relative efficiencies between signal and normalization channels,
of $0.71\pm0.01$, $0.74\pm0.02$, and $0.84\pm0.02$ for the $B^+$, $B^0$, and $B_s^0$ decays, respectively,
have uncertainties reaching 2.6\%, that arise from the finite size of the MC samples.

The predicted background values depend on the shape of the background
function. We evaluate the change in the background yield calculation when
using a sample that is similar to the one resulting from the optimal
selection, but has less stringent requirements on the three
optimization variables instead of the default
sample with poor vertex quality.
We compare two functional models for the background description, 
the default exponential plus a constant,
and a simpler linear extrapolation function.
We find a systematic uncertainty
on the background prediction reaching 10\% for the $B_s^0$ channel.

To calculate the systematic uncertainty on the relative branching
fraction, the relative efficiency uncertainty is summed in
quadrature with the uncertainty of the
$J/\psi\to\mu^+\mu^-$ branching ratio,
and with the systematic uncertainty 
on the number of signal candidates. 

To calculate the absolute branching fractions, we use the world average
branching fractions of the normalization channels~\cite{Yao:2006px}.
These branching fractions have uncertainties of 3.5\%, 4.5\%, and 35.5\%,
respectively, for $B^+$, $B^0$, and $B_s^0$,
which are added in quadrature to the systematic uncertainties on the
relative branching ratios.
A summary of the systematic
uncertainties is given in Table~\ref{tab:sys2}.

\begin{table}
\caption{Systematic uncertainties on the relative efficiency quoted in percent.}
\begin{center}
\begin{tabular}{lccc}
\hline\hline
Channel &$B^+$ & $B^0$ & $B_s$ \\\hline
Theory model                                            & 1.5 & 3.1 & 1.6 \\
$p_T(B)$ spectrum                                       & 0.6 & 1.3 & 1.4 \\
Trigger turn-on                                         & 1.3 & 1.3 & 1.2 \\
Low momentum hadrons                                    & 0.2 & 0.2 & 0.2 \\
$B_s^0$ decay width difference                          & $-$ & $-$ & 8.7 \\
Polarization                                            & $-$ & 0.6 & 0.1 \\
Norm. channel statistics                                & 1.3 & 2.1 & 5.1 \\
 $B^+\to J/\psi\pi^+$ contribution                      & 0.1 & $-$ & $-$ \\
MC statistics                                           & 1.6 & 2.6 & 2.2\\
\hline
Total &  2.9    &  5.0    & 10.6\\
\hline\hline
\end{tabular}
\end{center}
\label{tab:sys}
\end{table}

\begin{table}
\caption{Summary of systematic uncertainties quoted in percent.}
\begin{center}
\begin{tabular}{lccc}
\hline\hline
Channel     & $B^+$ & $B^0$ & $B_s$    \\\hline
Total rel. eff. uncertainty &  2.9    &  5.0    & 10.6\\
${\cal B}(J/\psi\to\mu^+\mu^-)$    &1.0     &1.0  &        1.0     \\
Background prediction     & 5.2  &3.1   & 10.2  \\
${\cal B}(B\to J/\psi h)$      & 3.5    & 4.5   &        35.5   \\
\hline\hline
\end{tabular}
\end{center}
\label{tab:sys2}
\end{table}

We calculate relative branching ratios using Eq.~(\ref{eqn:br}), and
find ${\cal B}(B^+\to\mu^+\mu^-K^+)/{\cal B}(B^+\to J/\psi K^+) =
(0.59\pm0.15\pm0.03)\times10^{-3}$, ${\cal
B}(B^0\to\mu^+\mu^-K^{*0})/{\cal B}(B^0\to J/\psi K^{*0}) =
(0.61\pm0.23\pm0.07)\times10^{-3}$, and ${\cal
B}(B_s^0\to\mu^+\mu^-\phi) /{\cal B}(B_s^0\to J/\psi \phi) =
(1.23\pm0.60\pm0.14)\times10^{-3}$, where the first uncertainty is
statistical, and the second is systematic.  We use the world-average
branching ratios of the $B^0$ and $B^+$ normalization channels~\cite{Yao:2006px},
 resulting in the following absolute branching fractions: ${\cal
B}(B^+\to\mu^+\mu^-K^+) =(0.59\pm0.15\pm0.04)\times10^{-6}$, and ${\cal
B}(B^0\to\mu^+\mu^-K^{*0})=(0.81\pm0.30\pm0.10)\times10^{-6}$.  To
obtain an absolute branching ratio for the
$B_s^0\to\mu^+\mu^-\phi$ decay, we use ${\cal B}(B_s^0\to
J/\psi\phi)=(1.38\pm 0.49)\times10^{-3}$, obtained from correcting the
CDF measurement~\cite{Abe} for the current value of $f_s/f_d$~\cite{Yao:2006px},
the $B_s^0$ to $B^0$ production ratio, resulting in ${\cal
B}(B_s^0\to\mu^+\mu^-\phi) =(1.70\pm0.82\pm0.64)\times10^{-6}$.
We find a significance of only 2.4 standard deviations for the $B_s^0\to\mu^+\mu^-\phi$ decay mode.
Therefore, we choose to set a limit on this decay.
We use a Bayesian integration assuming a flat prior~\cite{Heinrich:2004tj},
and find ${\cal B}(B_s^0\to\mu^+\mu^-\phi)/{\cal
B}(B_s^0\to J/\psi\phi)<2.6(2.3)\times 10^{-3}$ at the 95(90)\%
confidence level (C.L.). 
We also set an upper  limit on the
absolute branching ratio, taking into account the uncertainty
of the branching ratio of the normalization channel~\cite{D0norm}.
We assume that the prior probability
density function representing the uncertainty in the normalization
channel is a log normal distribution with a mean equal to the central
value, and a width parameter equal to the quoted uncertainty of 35.5\%.
We obtain ${\cal B}(B_s^0\to\mu^+\mu^-\phi)<6.0(5.0)\times 10^{-6}$ at 95(90)\% C.L.  The
main ingredients and the results are summarized in
Table~\ref{tab:signal}.

In conclusion, our measurements of the $B^+$ and $B^0$ rare decay modes are consistent with
the SM predictions, and with previous measurements~\cite{Aubert:2006vb,Ishikawa:2003cp}.
The relative limit on $B_s^0\to\mu^+\mu^-\phi$ is consistent with the SM predictions, 
and is the most stringent to date.

\begin{table*}
\caption{Summary of the main ingredients and results of this analysis.
Where one uncertainty is quoted, the uncertainty is of statistical nature.
Where two uncertainties are quoted, the first is statistical, and the second is systematic.}
\begin{center}
\begin{tabular}{lccc}
\hline\hline
Decay mode                  & $B^+ \to \mu^+\mu^- K^+$  & $B^0\to\mu^+\mu^- K^{*0}$  & $B_s^0\to\mu^+\mu^- \phi$   \\\hline 
$N_{obs}$      & 90           & 35 	          & 11                    \\ 
$N_{bkg}$              & $45.3\pm5.8$	     & $16.5\pm3.6$         & $3.5\pm1.5$          \\ 
$N_{sig}$      & $44.7\pm5.8$           & $18.5\pm3.6$ 	          & $7.5\pm1.5$                    \\ 
Gaussian significance & 4.5\,$\sigma$     & 2.9\,$\sigma$                & 2.4\,$\sigma$          \\ 
$N_{J/\psi h}$            & $6361\pm82$          & $2423\pm52$       & $431\pm22$        \\
$\epsilon_{\mu^+\mu^-h}/\epsilon_{J/\psi h}$ & $0.71\pm0.01$        & $0.74\pm0.02$      & $0.84\pm0.02$        \\
Rel ${\cal B} \times 10^3$& $0.59\pm0.15\pm0.03$  &$0.61\pm0.23\pm0.07$  & $1.23\pm0.60\pm0.14$ \\ 
Abs ${\cal B} \times 10^6$& $0.59\pm0.15\pm0.04$  & $0.81\pm0.30\pm0.10$  & $1.70\pm0.82\pm0.64$ \\ 
Rel ${\cal B}$ 95(90)\%C.L. limit $\times 10^3$     &   ---                &---                   &2.6(2.3)                \\ 
Abs ${\cal B}$ 95(90)\%C.L. limit $\times 10^6$     &   ---                & ---                  &6.0(5.0)                \\
\hline\hline 
\end{tabular}
\end{center}
\label{tab:signal}
\end{table*}

\begin{acknowledgments}
We thank the Fermilab staff and the technical staffs of the
participating institutions for their vital contributions. This work
was supported by the U.S. Department of Energy and National Science
Foundation; the Italian Istituto Nazionale di Fisica Nucleare; the
Ministry of Education, Culture, Sports, Science and Technology of
Japan; the Natural Sciences and Engineering Research Council of
Canada; the National Science Council of the Republic of China; the
Swiss National Science Foundation; the A.P. Sloan Foundation; the
Bundesministerium f\"ur Bildung und Forschung, Germany; the Korean
Science and Engineering Foundation and the Korean Research Foundation;
the Science and Technology Facilities Council and the Royal Society,
UK; the Institut National de Physique Nucleaire et Physique des
Particules/CNRS; the Russian Foundation for Basic Research; the
Comisi\'on Interministerial de Ciencia y Tecnolog\'{\i}a, Spain; the
European Community's Human Potential Programme; the Slovak R\&D Agency;
and the Academy of Finland.
\end{acknowledgments}

\end{document}

%% file: Jan_2008_Authors_Visitors1_oldeman_tesarek.tex
\affiliation{Institute of Physics, Academia Sinica, Taipei, Taiwan 11529, Republic of China} 
\affiliation{Argonne National Laboratory, Argonne, Illinois 60439} 
\affiliation{University of Athens, 157 71 Athens, Greece} 
\affiliation{Institut de Fisica d'Altes Energies, Universitat Autonoma de Barcelona, E-08193, Bellaterra (Barcelona), Spain} 
\affiliation{Baylor University, Waco, Texas  76798} 
\affiliation{Istituto Nazionale di Fisica Nucleare Bologna, $^t$University of Bologna, I-40127 Bologna, Italy} 
\affiliation{Brandeis University, Waltham, Massachusetts 02254} 
\affiliation{University of California, Davis, Davis, California  95616} 
\affiliation{University of California, Los Angeles, Los Angeles, California  90024} 
\affiliation{University of California, San Diego, La Jolla, California  92093} 
\affiliation{University of California, Santa Barbara, Santa Barbara, California 93106} 
\affiliation{Instituto de Fisica de Cantabria, CSIC-University of Cantabria, 39005 Santander, Spain} 
\affiliation{Carnegie Mellon University, Pittsburgh, PA  15213} 
\affiliation{Enrico Fermi Institute, University of Chicago, Chicago, Illinois 60637} 
\affiliation{Comenius University, 842 48 Bratislava, Slovakia; Institute of Experimental Physics, 040 01 Kosice, Slovakia} 
\affiliation{Joint Institute for Nuclear Research, RU-141980 Dubna, Russia} 
\affiliation{Duke University, Durham, North Carolina  27708} 
\affiliation{Fermi National Accelerator Laboratory, Batavia, Illinois 60510} 
\affiliation{University of Florida, Gainesville, Florida  32611} 
\affiliation{Laboratori Nazionali di Frascati, Istituto Nazionale di Fisica Nucleare, I-00044 Frascati, Italy} 
\affiliation{University of Geneva, CH-1211 Geneva 4, Switzerland} 
\affiliation{Glasgow University, Glasgow G12 8QQ, United Kingdom} 
\affiliation{Harvard University, Cambridge, Massachusetts 02138} 
\affiliation{Division of High Energy Physics, Department of Physics, University of Helsinki and Helsinki Institute of Physics, FIN-00014, Helsinki, Finland} 
\affiliation{University of Illinois, Urbana, Illinois 61801} 
\affiliation{The Johns Hopkins University, Baltimore, Maryland 21218} 
\affiliation{Institut f\"{u}r Experimentelle Kernphysik, Universit\"{a}t Karlsruhe, 76128 Karlsruhe, Germany} 
\affiliation{Center for High Energy Physics: Kyungpook National University, Daegu 702-701, Korea; Seoul National University, Seoul 151-742, Korea; Sungkyunkwan University, Suwon 440-746, Korea; Korea Institute of Science and Technology Information, Daejeon, 305-806, Korea; Chonnam National University, Gwangju, 500-757, Korea} 
\affiliation{Ernest Orlando Lawrence Berkeley National Laboratory, Berkeley, California 94720} 
\affiliation{University of Liverpool, Liverpool L69 7ZE, United Kingdom} 
\affiliation{University College London, London WC1E 6BT, United Kingdom} 
\affiliation{Centro de Investigaciones Energeticas Medioambientales y Tecnologicas, E-28040 Madrid, Spain} 
\affiliation{Massachusetts Institute of Technology, Cambridge, Massachusetts  02139} 
\affiliation{Institute of Particle Physics: McGill University, Montr\'{e}al, Canada H3A~2T8; and University of Toronto, Toronto, Canada M5S~1A7} 
\affiliation{University of Michigan, Ann Arbor, Michigan 48109} 
\affiliation{Michigan State University, East Lansing, Michigan  48824}
\affiliation{Institution for Theoretical and Experimental Physics, ITEP, Moscow 117259, Russia} 
\affiliation{University of New Mexico, Albuquerque, New Mexico 87131} 
\affiliation{Northwestern University, Evanston, Illinois  60208} 
\affiliation{The Ohio State University, Columbus, Ohio  43210} 
\affiliation{Okayama University, Okayama 700-8530, Japan} 
\affiliation{Osaka City University, Osaka 588, Japan} 
\affiliation{University of Oxford, Oxford OX1 3RH, United Kingdom} 
\affiliation{Istituto Nazionale di Fisica Nucleare, Sezione di Padova-Trento, $^u$University of Padova, I-35131 Padova, Italy} 
\affiliation{LPNHE, Universite Pierre et Marie Curie/IN2P3-CNRS, UMR7585, Paris, F-75252 France} 
\affiliation{University of Pennsylvania, Philadelphia, Pennsylvania 19104} 
\affiliation{Istituto Nazionale di Fisica Nucleare Pisa, $^q$University of Pisa, $^r$University of Siena and $^s$Scuola Normale Superiore, I-56127 Pisa, Italy} 
\affiliation{University of Pittsburgh, Pittsburgh, Pennsylvania 15260} 
\affiliation{Purdue University, West Lafayette, Indiana 47907} 
\affiliation{University of Rochester, Rochester, New York 14627} 
\affiliation{The Rockefeller University, New York, New York 10021} 

\affiliation{Istituto Nazionale di Fisica Nucleare, Sezione di Roma 1, $^v$Sapienza Universit\`{a} di Roma, I-00185 Roma, Italy} 

\affiliation{Rutgers University, Piscataway, New Jersey 08855} 
\affiliation{Texas A\&M University, College Station, Texas 77843} 
\affiliation{Istituto Nazionale di Fisica Nucleare Trieste/\ Udine, $^w$University of Trieste/\ Udine, Italy} 
\affiliation{University of Tsukuba, Tsukuba, Ibaraki 305, Japan} 
\affiliation{Tufts University, Medford, Massachusetts 02155} 
\affiliation{Waseda University, Tokyo 169, Japan} 
\affiliation{Wayne State University, Detroit, Michigan  48201} 
\affiliation{University of Wisconsin, Madison, Wisconsin 53706} 
\affiliation{Yale University, New Haven, Connecticut 06520} 
\author{T.~Aaltonen}
\affiliation{Division of High Energy Physics, Department of Physics, University of Helsinki and Helsinki Institute of Physics, FIN-00014, Helsinki, Finland}
\author{J.~Adelman}
\affiliation{Enrico Fermi Institute, University of Chicago, Chicago, Illinois 60637}
\author{T.~Akimoto}
\affiliation{University of Tsukuba, Tsukuba, Ibaraki 305, Japan}
\author{M.G.~Albrow}
\affiliation{Fermi National Accelerator Laboratory, Batavia, Illinois 60510}
\author{B.~\'{A}lvarez~Gonz\'{a}lez}
\affiliation{Instituto de Fisica de Cantabria, CSIC-University of Cantabria, 39005 Santander, Spain}
\author{S.~Amerio$^u$}
\affiliation{Istituto Nazionale di Fisica Nucleare, Sezione di Padova-Trento, $^u$University of Padova, I-35131 Padova, Italy} 

\author{D.~Amidei}
\affiliation{University of Michigan, Ann Arbor, Michigan 48109}
\author{A.~Anastassov}
\affiliation{Northwestern University, Evanston, Illinois  60208}
\author{A.~Annovi}
\affiliation{Laboratori Nazionali di Frascati, Istituto Nazionale di Fisica Nucleare, I-00044 Frascati, Italy}
\author{J.~Antos}
\affiliation{Comenius University, 842 48 Bratislava, Slovakia; Institute of Experimental Physics, 040 01 Kosice, Slovakia}
% ...

\collaboration{CDF Collaboration\footnote{With visitors from $^a$Universiteit Antwerpen, B-2610 Antwerp, Belgium, 
$^b$Chinese Academy of Sciences, Beijing 100864, China, 
$^c$University of Bristol, Bristol BS8 1TL, United Kingdom, 
$^d$University of California Irvine, Irvine, CA  92697, 
$^e$University of California Santa Cruz, Santa Cruz, CA  95064, 
$^f$Cornell University, Ithaca, NY  14853, 
$^g$University of Cyprus, Nicosia CY-1678, Cyprus, 
$^h$University College Dublin, Dublin 4, Ireland, 
$^i$University of Edinburgh, Edinburgh EH9 3JZ, United Kingdom, 
$^j$Universidad Iberoamericana, Mexico D.F., Mexico, 
$^k$University of Manchester, Manchester M13 9PL, England, 
$^l$Nagasaki Institute of Applied Science, Nagasaki, Japan, 
$^m$University de Oviedo, E-33007 Oviedo, Spain, 
$^n$Queen Mary, University of London, London, E1 4NS, England, 
$^o$Texas Tech University, Lubbock, TX  79409, 
$^p$IFIC(CSIC-Universitat de Valencia), 46071 Valencia, Spain,
$^x$Royal Society of Edinburgh/Scottish Executive Support Research Fellow, 
}}
\noaffiliation

%% file: bmumuhdraft2_2.bbl
\begin{thebibliography}{99}
\bibitem{eta}
The CDF reference frame uses cylindrical coordinates, where
$\theta$ and $\phi$ are the polar and azimuthal angles with respect to
the proton beam.  Pseudorapidity $\eta$ is defined as $-\ln (\tan(\theta/2))$.
\end{thebibliography}
